\documentclass[prd,twocolumn,showpacs]{revtex4}
\usepackage{amsmath,amssymb,mathtools,amscd}
\usepackage{epsfig,amsmath,amssymb,mathtools,amscd}
\usepackage[active]{srcltx}
\usepackage{dsfont}
\usepackage{subfigure}
\usepackage{enumerate}

\mathchardef\minus="002D

\def\<{\langle}\def\>{\rangle}

\def\BraKet#1#2{\left\langle #1\middle | #2 \right\rangle} 
 \def\ket#1{| #1 \rangle} 
\def\bra#1{\langle #1 |} 
\def\ketbra#1#2{| #1 \rangle \langle#2 |}

\def\Z{\mathbb Z}

\def\L2{{\mathcal L}_2}

\def\s{\bar{s}}

\def\d#1 {\mathop{\!\! \mathrm{d}#1}\,}
\def\df#1#2 {\!\!\frac{\mathop{\mathrm{d}#1}}{#2}\,}

\newcommand{\dd}{\mathop{\!\! \mathrm{d}}}

\begin{document}
\title{The Dirac Quantum Cellular Automaton in one dimension: Zitterbewegung and scattering from
  potential} \author{Alessandro 
  \surname{Bisio}} \email[]{alessandro.bisio@unipv.it}
\affiliation{Dipartimento di Fisica dell'Universit\`a di Pavia, via
  Bassi 6, 27100 Pavia} \affiliation{Istituto Nazionale di Fisica
  Nucleare, Gruppo IV, via Bassi 6, 27100 Pavia} \author{Giacomo Mauro
  \surname{D'Ariano}} \email[]{dariano@unipv.it}
\affiliation{Dipartimento di Fisica dell'Universit\`a di Pavia, via
  Bassi 6, 27100 Pavia} \affiliation{Istituto Nazionale di Fisica
  Nucleare, Gruppo IV, via Bassi 6, 27100 Pavia} \author{Alessandro
  \surname{Tosini}}\email[]{alessandro.tosini@unipv.it}
\affiliation{Dipartimento di Fisica dell'Universit\`a di Pavia, via
  Bassi 6, 27100 Pavia} \affiliation{Istituto Nazionale di Fisica
  Nucleare, Gruppo IV, via Bassi 6, 27100 Pavia}
\begin{abstract}
  We study the dynamical behaviour of the quantum cellular automaton
  of Refs.  \cite{darianopla,BDTqcaI}, which reproduces the Dirac
  dynamics in the limit of small wavevectors and masses.  We present
  analytical evaluations along with computer simulations, showing how
  the automaton exhibits typical Dirac dynamical features, as the
  Zitterbewegung and the scattering behaviour from potential that
  gives rise to the so-called Klein paradox. The motivation is to show
  concretely how pure processing of quantum information can lead to
  particle mechanics as an emergent feature, an issue that has been
  the focus of solid-state, optical and atomic-physics
  quantum simulator.
\end{abstract}
\maketitle  
\section{Introduction}

The idea of reproducing the evolution of a macroscopic system starting from a simple rule of local
interaction among its elementary constituents was first formalized in the pioneering von Neumann's
paper \cite{neumann1966theory} with the notion of \emph{Cellular Automaton}. The automaton is a
regular lattice of cells with a finite number of states, equipped with a rule that updates the cell
states from time $t$ to time $t+1$.  Such rule must be \emph{local}, namely the state of the $x$
cell at $t+1$ depends only on the states of a finite number of neighboring cells at $t$. Cellular
automata have been a popular topic for many years, as a new paradigm for complex systems, and many
books have been devoted to the subject (see eg. Refs. \cite{wolfram2002new,toffoli1987cellular}).
One of the reasons of its first success, which eventually has become its own weakness, is the
chaotic behaviour of the automaton dynamics \cite{sep-cellular-automata}. 

Differently from classical cellular automata, \emph{Quantum Cellular Automata} (QCA) exhibit a less
chaotic behaviour, which makes them predictable for large number of steps \cite{ambainis2001one}.
Here the cells are finite-dimensional quantum systems interacting locally and unitarily. Being
locality of interactions an essential ingredient of any physical evolution, QCA have been
considered already by Feynman as candidates for simulating physics
\cite{feynman1965quantum,feynman1982simulating}.  More recently QCA earned interest in the quantum
information community leading to many results on its mathematical theory
\cite{schumacher2004reversible,arrighi2011unitarity,gross2012index}, and on their general dynamical
features
\cite{ambainis2001one,knight2004propagating,valcarcel2010tailoring,ahlbrecht2011asymptotic,reitzner2011quantum}.
In quantum field theory, after the first appearance of a prototype of QCA in the Feynman chessboard
\cite{feynman1965quantum} for solving the path-integral for the Dirac field, a similar framework has
appeared in the work of Nakamura \cite{nakamura1991nonstandard} motivated by a rigorous formulation
of the Feynman path integral, and later in the seminal work of Bialynicki-Birula
\cite{bialynicki1994weyl}, as a lattice theory for Weyl, Dirac, and Maxwell fields. Then the
possibility of using automata for describing the evolution of relativistic fields emerged in the
context of lattice-gas simulations, especially in the work of Meyer \cite{meyer1996quantum}, where a
notion of ``field automaton'' first appeared, and in the papers of Yepez \cite{Yepez:2006p4406}.

More recently QCA have been considered for extending quantum field theory \cite{darianopla} to the
Plank scale. Similar to lattice-gas theories, here the quantum cell corresponds to the evaluation
$\psi(x)$ of a quantum field on the site $x$ of a lattice, with the dynamics updated in discrete
time steps by a local unitary evolution. However, differently from lattice-gas theory, here the
continuum limit is not taken, whereas, instead, the asymptotic large-scale (Fermi) evolution is
considered.  The main difference is then that Lorentz covariance holds exactly in the relativistic
limit of small momentum, whereas generally it is distorted, in a fashion analogous to Refs.
\cite{magueijo2003generalized,amelino2001planck1,amelino2001planck}. In this context the one
dimensional Dirac automaton has been derived from symmetry principles for the QCA \cite{BDTqcaI}
showing how the usual Dirac dynamics emerges at the Fermi scale, though relativistic covariance and
other symmetries are violated at the Planck/ultrarelativistic scale.

In the present paper we analyze in detail the one-particle sector of the automaton of Refs.
\cite{darianopla,BDTqcaI}. Here, particle states are ``smooth'' states peaked around a momentum
eigenstate of the QCA. We will consider dynamical quantities as the particle position, momentum and
velocity, along with their evolution both in the free case and in the presence of a potential,
recovering typical features of Dirac quantum field evolution--as {\em Zitterbewegung} and {\em
  Klein paradox}--from the pure quantum information processing of the QCA. Recently there has
been a renewed interest in Dirac features in solid-state and atomic physics, which provide a
physical hardware to simulate the dynamics. Zitterbewegung can be seen in the response of electrons
to external fields \cite{huang1952zitterbewegung} and can appear for nonrelativistic particles in a
crystal \cite{cannata1991effects,ferrari1990nonrelativistic,cannata1990dirac}, quasiparticles in
superconductors \cite{lurie1970zitterbewegung} and systems with spin-orbit coupling
\cite{PhysRevLett.95.187203,PhysRevLett.99.076603}. Proving that the oscillation behavior is not
unique to Dirac electrons, but rather is a generic feature of spinor systems with linear dispersion
relations, these works opened the way for possible simulation of Zitterbewegung using for example
trapped ions \cite{lamata2007dirac,gerritsma2010quantum}, two-band crystalline structure such as
graphene \cite{cserti2006unified,rusin2007transient} or semiconductors
\cite{schliemann2005zitterbewegung,zawadzki2005zitterbewegung,zawadzki2010nature,geim2007rise,zawadzki2011zitterbewegung},
ultra cold atoms \cite{vaishnav2008obserVing}, and finally photonic crystals
\cite{PhysRevLett.100.113903}. On the other hand, the Klein paradox (tunneling of relativistic
particles) provides insight in the mechanics of relativistic particles propagating through potential
barriers, along with vacuum polarization effects, and has been a focus in the hot topic of graphene
as a simulator for Dirac equation, as in Ref. \cite{katsnelson2007graphene}, and
\cite{gerritsma2010quantum} for trapped ions. Recently also microfabricated optical waveguide
circuits have become an alternative physical simulator for particle dynamics \cite{sansoni2012two}.

After reviewing the Dirac QCA in 1d in Section \ref{s:DiracQCA}, in Section \ref{sec:zitterbewegung}
we present the evolution of position and momentum operators for the automaton, showing the
Zitterbewegung behaviour produced by the interference between positive and negative frequencies. 
In Section \ref{s:barrier} we modify the QCA in order to insert a potential in the free evolution, and
show the automaton dynamics in the presence of a barrier for one particle states.

We end the paper with a summary and some concluding remarks in Section\ref{s:concl}.

\section{The Dirac Automaton}\label{s:DiracQCA}
The quantum automaton corresponding to the Dirac equation in 1d, first
introduced in \cite{darianopla}, has been derived from the discrete automaton symmetries of parity and
time-reversal in Ref.  \cite{BDTqcaI}, where also the Dirac equation has been recovered as
the large-scale relativistic limit of the automaton. The cell of the quantum automaton is given by the
evaluation $\psi(x)$ of the two-component field operator $\psi$, and the unitary evolution of
one step of the automaton is given by
\begin{align}
\psi(x)\to U\psi(x),\quad
\psi(x) := 
\left(\begin{array}{c}
    \psi_r(x) \\
\psi_l(x)
  \end{array} 
\right)\quad 
\end{align}
where $\psi_l$ and $\psi_r$ denote the {\em left} and {\em right} mode of the field, whereas the unitary
matrix $U$ is given by
\begin{equation}\label{U}
U=\begin{pmatrix}
  n S &-im\\
  -im & n S^\dag
\end{pmatrix},\\
\quad n^2+m^2=1,
\end{equation}
with $S$ denoting the shift operator $S f(x) = f(x+1)$.  The constants $n$ and $m$ in the last
equation can be chosen positive. As shown in Refs. \cite{darianopla,BDTqcaI}, the parameter $m$
plays the role of an a-dimensional inertial mass, and is bounded by unit. We remark that the
automaton description is completely a-dimensional, and a conversion to the usual physical dimensions
needs a length, a time and a mass, which one can take as the Planck length $\ell_P$, the Planck time
$\tau_P$, and the Planck mass $m_P$, the latter playing the role of the bound for the inertial mass.
The maximal speed of propagation of information is one cell per step ($c=\ell_P/\tau_P$ in
dimensional units, corresponding to the speed of light). The quantum field can be taken generally as
Fermionic, Bosonic, or even Anyonic. However, in the present case it will not be relevant, since we
will consider only single-particle states, which span the Hilbert space $\mathbb{C}^2\otimes
l_2(\Z)$, and for which we will use the factorized orthonormal basis  $\ket{s}\ket{x}$, where for
$\ket{s}$ we consider the canonical basis corresponding to $s=l,r$. These states can be also
obtained as $\psi^\dag_s(x)\ket{\Omega}$ upon introducing a vacuum $|\Omega\>$ which is annihilated
by the field operator, and invariant under the automaton evolution. Similarly also $N$-particle
states with $N>1$ can be obtained by acting with products of $N$ evaluations of the field operator,
building up the Fock space in the usual way. Notice that the evolution of the field is restricted to
be linear, and there exists a unitary operator $U$ such that the field evolution is given by
$V\psi_s(x)V^\dag=U\psi_s(x)$, with $V|\Omega\>=|\Omega\>$, whereas for product of field
evaluations the evolution is given by tensor powers of $U$ as $V\psi_{s_1}(x_1) \ldots
\psi_{s_N}(x_N) V^\dag=U^{\otimes N}\psi_{s_1}(x_1)\otimes
\ldots\otimes\psi_{s_N}(x_N)$.  

In the $\ket{s}\ket{x}$ representation the unitary matrix $U$ can be written as follows 

\begin{align}
  U := \sum_x
\begin{pmatrix}
   n \ketbra{x-1}{x} &   -im \ketbra{x}{x} \\
  -im \ketbra{x}{x} &   n \ketbra{x+1}{x} 
  \end{pmatrix},
\end{align}
describing a {\em Quantum Walk} on the Hilbert space
$\mathbb{C}^2\otimes l_2(\Z)$ \cite{ambainis2001one}.

Tanks to the translational invariance of $U$, it is convenient to move to the momentum
representation
\begin{align}\label{vk}
  \ket{\psi_s}\ket{k} :=\frac{1}{\sqrt{2\pi}}  \sum_{x} e^{-ikx}  \ket{\psi_s}\ket{x},\quad k\in[-\pi,\pi],
\end{align}
and $U$ becomes
\begin{align}\label{eq:automaton-Uk}
U=
\int_{\minus \pi}^{\pi} \d k U(k)\otimes\ketbra{k}{k},\:\, {U}(k)=
\begin{pmatrix}
  n e^{ik} & -i m\\
  -i m & n e^{-ik}
\end{pmatrix}.
\end{align}
Notice that discreteness bound momenta to the Brillouin zone, as in solid-state theory.
By diagonalizing the unitary matrix ${U}(k)$ 
\begin{align}\label{eq:eigenstates}
&U(k) \ket{s}_k = e^{-is \omega(k)}\ket{s}_k,\qquad \omega(k) = \arccos(n \cos k)\\\nonumber
&\ket{s}_k:=\tfrac{1}{\sqrt{2}}
\begin{bmatrix}
\sqrt{1-sv(k)}\\s\sqrt{1+sv(k)}
\end{bmatrix}, \quad s=\pm,\quad  v(k) := \partial_k \omega(k)
\end{align}
it is easy to evaluate the logarithm of $U$ ($e^{-i H}:= U$) as follows
\begin{align}\label{eq:hamiltonian}
H = \int_{-\pi}^{\pi} \d k H(k)&\otimes\ketbra{k}{k},\\\nonumber
H(k) &=\omega(k)\left( \ket{+}_k\bra{+}_k -
\ket{-}_k\bra{-}_k\right) \\\nonumber
&=\mathrm{sinc^{-1}}{\omega(k)} (-n\sin k\, \sigma_3+m\sigma_1),
\end{align}
where $\sigma_i$ $i=1,2,3$ denote the usual Pauli matrices. 

\begin{figure}[t!]
\includegraphics[width=.21\textwidth]{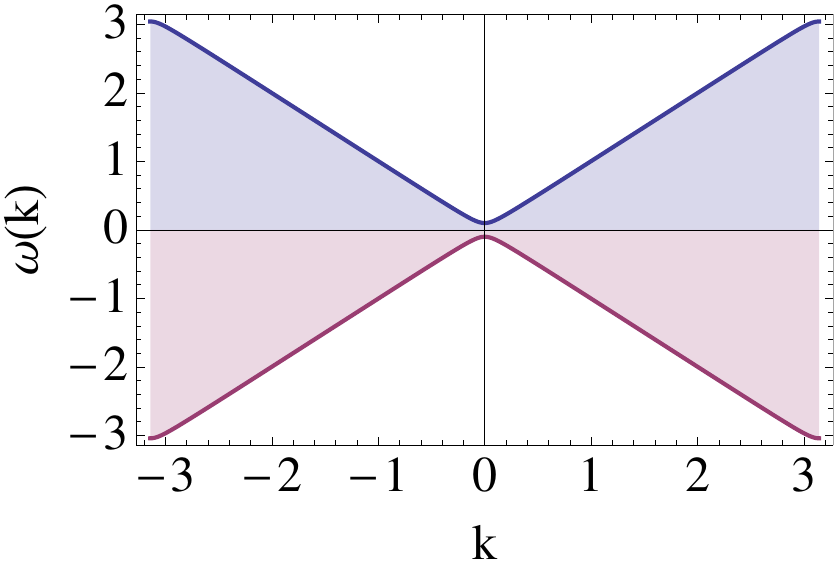}\quad
\includegraphics[width=.21\textwidth]{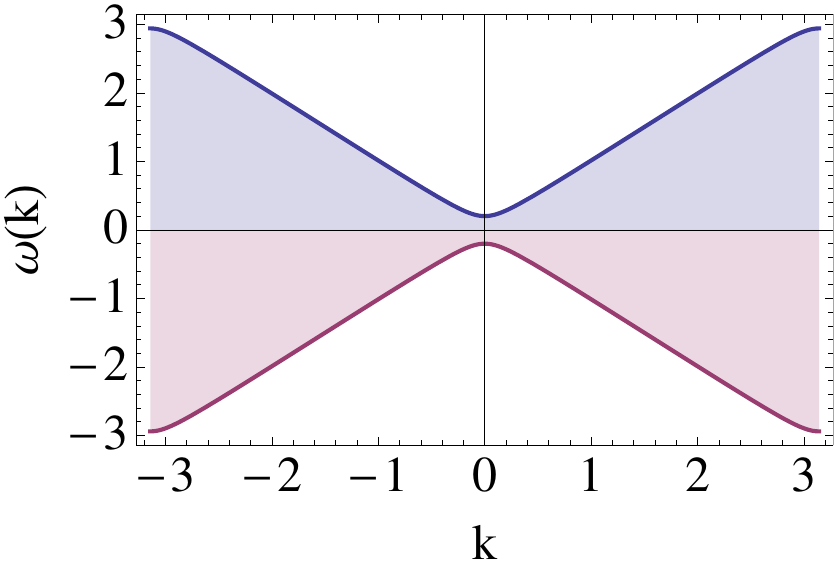}
\includegraphics[width=.21\textwidth]{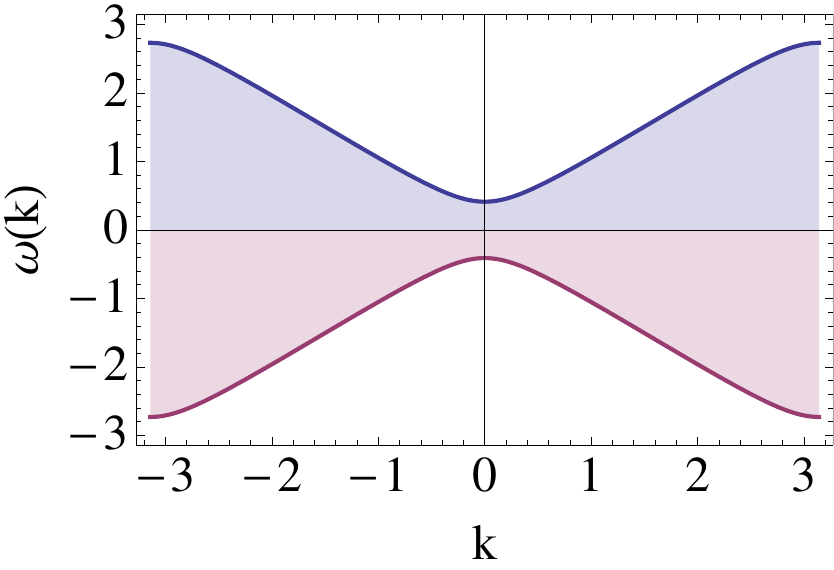}\quad
\includegraphics[width=.21\textwidth]{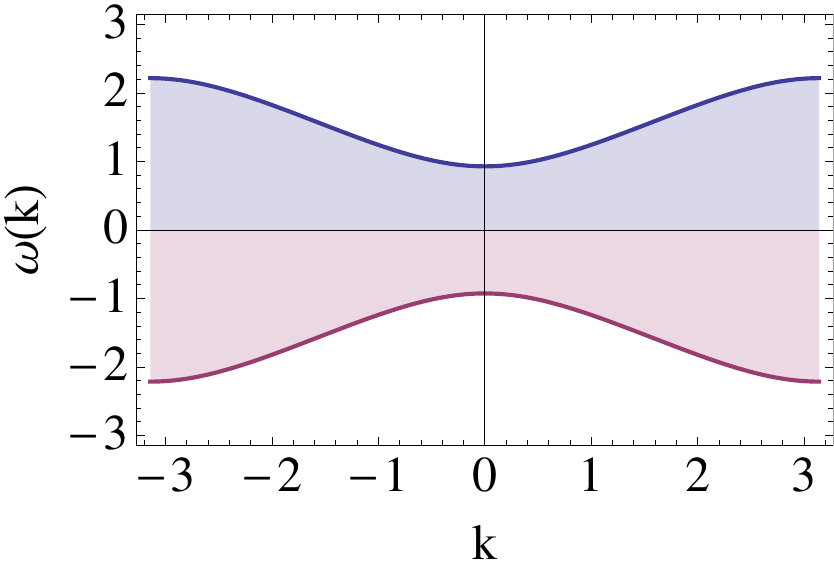}
\caption{The Dirac automaton dispersion relation in Eq. \eqref{eq:eigenstates}
  for four different values of the mass: $m=0.1,\,0.2,\,0.4,\,0.8$.}\label{fig:disp}
\end{figure}

The function $\omega(k)$ is the dispersion relation of the automaton, which recovers the usual Dirac
one $\omega(k)=\sqrt{k^2+m^2}$ in the limit $k,m\ll 1$ and $k/m\gg 1$ as shown in \cite{BDTqcaI}.
This is also clear in Fig. \ref{fig:disp} where the dispersion relation as a function of $k$ is
reported for four different values of the mass. The derivative $v(k)$ in Eq.  (\ref{eq:eigenstates})
is then the group velocity of the wavepacket.  The $s=+1$ eigenvalues correspond to
positive-energy particle states, whereas the negative $s=-1$ eigenvalues correspond to
negative-energy anti-particle states. 

Notice that the operator $H$ regarded as an Hamiltonian would interpolate the evolution to continuous
time as $U(t)\equiv U^t$, which, however, in this context should be
considered unphysical
\footnote{The interactions between cells in the interpolation interval would be nonlocal, and, in
  addition, the Hamiltonian would involve distant cells.}  

 In the following sections we will analyze two typical aspects of the Dirac-field dynamics, namely
 the Zitterbewegung and the Klein paradox. 

\section{Position and Momentum Operators and Zitterbewegung}\label{sec:zitterbewegung}

The QCA (\ref{U}) describes very precisely the Dirac field dynamics for customary relativistic
wavevectors and energies (consider that e.g. a ultra-high-energy cosmic ray has $k\simeq 10^{-8}$)
\cite{BDTqcaI}. In this section we will show how efficiently it reproduces a typical feature of the
one-particle Dirac dynamics, namely the Zitterbewegung.
 
The Zitterbewegung was first recognized by Schr\"odinger in 1930 \cite{schrodinger1930kraftefreie}
who noticed that in the Dirac equation describing the free relativistic electron the velocity
operator does not commute with the Dirac Hamiltonian: the evolution of the position operator, in
addition to the classical motion shows a very fast periodic oscillation with frequency $2mc^2$ and
amplitude equal to the Compton wavelength $\hbar/mc$ with $m$ the rest mass of the
relativistic particle. This jittering motion first encountered in the Dirac theory of the electron
was then shown \cite{huang1952zitterbewegung} to arise from the interference of states corresponding
to the positive and negative energies resulting from the Dirac equation with the trembling
disappearing with time \cite{lock1979zitterbewegung} for a wavepacked particle state.
Zitterbewegung oscillations cannot be directly observed by current experimental techniques for a
Dirac electron since the amplitude should by very small $\approx 10^{-12}$ m.  However,
it can be seen in a number of solid-state, atomic-physics, photonic-cristal and optical waveguide
simulators, as quoted in the introduction.

The ``position'' operator $X$ providing the representation $|x\>$ (i.e. such that
$X|s\>|x\>=x|s\>|x\>$) is defined as follows
\begin{align}
X=\sum_{x\in\mathbb{Z}}x(I\otimes\ketbra{x}{x}).
\end{align}
Generally $X$ provides the average location of a wavepacket in terms of $\<\psi|X|\psi\>$. The
conjugated ``momentum'' operator is given by
\begin{align}
P=\int_{-\pi}^{\pi}\;\df{k}{2\pi} k(I\otimes\ketbra{k}{k}).
\end{align}

One can verify that $X$ and $P$ obey the usual canonical commutation rule $[X,P]=i$. In the
following it will be convenient to work with the continuous time $t$ interpolating exactly the
discrete automaton evolution, namely $U^t$. However, all numerical results will be given only for
discrete $t$, namely for repeated applications of the automaton unitary $U$ in Eq.  (\ref{U}).

The time evolution of the position operator ${X}(t)={U}^{ - t} {X}{U}^{t}$ can be more easily
computed by integrating the differential equation $A(t) = [H,[H,X(t)]]$ where $H$ was defined in Eq.
\eqref{eq:hamiltonian}. We have
\begin{align}\nonumber
 &A(t) = \int_{-\pi}^{\pi} \d k  A(k,t) \otimes\ketbra{k}{k} \quad A(k,t) =
 e^{2i{H(k)}t}{A}(k)\\
&A(k) = -\frac{2\omega^2}{\sin^2\omega}nm\cos{k}\,
\sigma_2
\end{align}
which leads to
\begin{align}\label{e:Zx}
  {X}(t)={X}(0)+{V} t+{Z}_{{X}}(t)-{Z}_{{X}}(0)\\
  {V}(k)=-v(k)^2\sigma_3+v(k)\sqrt{1-v(k)^2}\sigma_1\\
  {Z}_{{X}}(k,t)=-\frac{1}{4}{H}^{-2}(k){A}(k,t)
\end{align}
where $V$ is the classical component of the velocity operator which, in the base diagonalizing the
Hamiltonian \eqref{eq:hamiltonian}, is $V(k)=v(k)\sigma_3$ and is proportional to the group velocity
$v(k)$. Since a generic one-particle state $\ket{\psi}$ is a superposition of a positive and a
negative energy state, i.e.  $\ket{\psi_+} + \ket{\psi_-}$, the evolution of the mean value of the
position operator $X(t)$, can be written as
\begin{align}\nonumber
x_\psi(t) :=  \bra{\psi} X(t) \ket{\psi} =
x_{\psi}^+(t) + x_{\psi}^-(t) + x_{\psi}^{\rm{int}}(t)\\\nonumber
x_{\psi}^{\pm}(t) : = \bra{\psi _\pm} X(0) + Vt \ket{\psi_\pm}\\\label{e:zitterbewegung}
x_{\psi}^{\rm{int}}(t) :=  
2 \Re [\bra{\psi_+} X(0) - {Z}_{{X}}(0) + {Z}_{{X}}(t) \ket{\psi_-}]
\end{align}
where $\Re$ denotes the real part.  The interference between positive
and negative frequency is responsible of the $x_{\psi}^{\rm{int}}(t)$.
The magnitude of $x_{\psi}^{\rm{int}}(t)$ is bounded by $1/m$ (see
appendix \ref{a:zitterbewegung}) which in the usual dimensional units
correspond to the Compton wavelength $\hbar/ m c$.  Moreover the
stationary phase approximation shows that for $t \to \infty$ the term
$2 \Re [\bra{\psi_+} {Z}_{{X}}(t) \ket{\psi_-}]$, which is responsible
of the oscillation, goes to $0$ as $1/\sqrt{t}$ (see appendix
\ref{a:zitterbewegung}) and only the shift
contribution coming from $2 \Re [\bra{\psi_+} X(0) - {Z}_{{X}}(0)
\ket{\psi_-}]$ survives.  These results show that
$x_{\psi}^{\rm{int}}(t)$ is the automaton analogue of the so-called
Zitterbewegung. As already noticed in the introduction this phenomenon
was never observed for a free relativistic electron because of the
small value of $x_{\psi}^{\rm{int}}(t)$ which is bounded by the
electron Compton wave length $10^{-12}$ m in natural units.  The
results of this section are in agreement with the one for the
Hadamard walk \cite{kurzynski2008relativistic}.

\begin{figure}[t!]
\includegraphics[width=.23\textwidth]{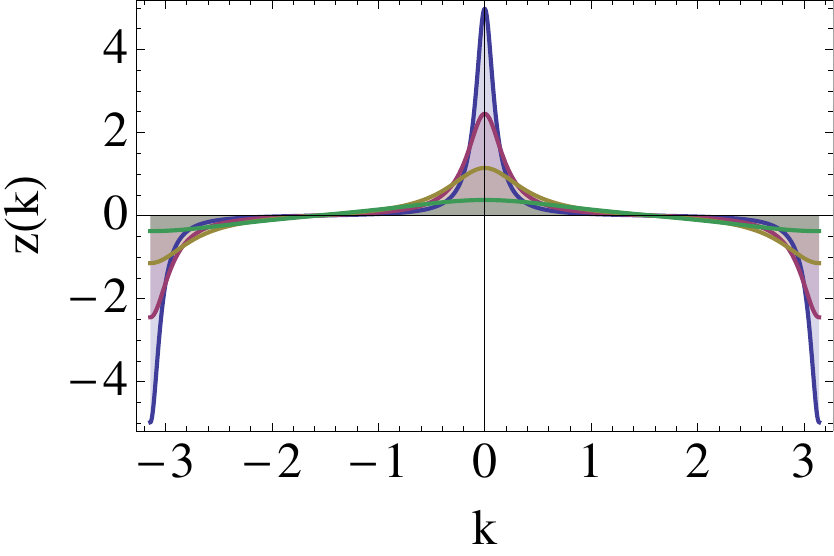}\quad
\includegraphics[width=.23\textwidth]{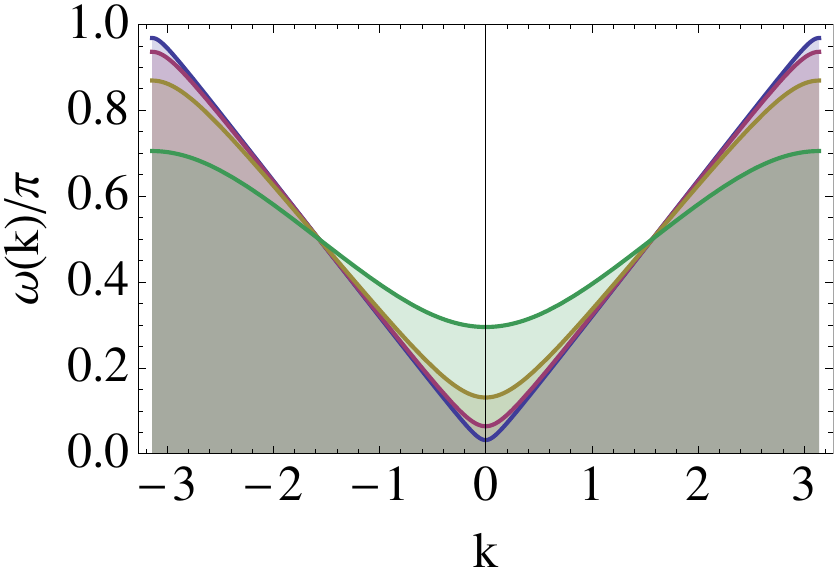}
\caption{(Colors online) Plots of $z(k)$ (left) and $\omega(k)/\pi$ (right) related to the
  oscillation amplitude and frequency of the position expectation value in Eq. \eqref{e:Zx}. In both
  cases the plots are reported for different values of the mass ($m=0.1,\,0.2,\,0.4,\,0.8$ from the
  top in the figure on the left and from the bottom in the figure on the right).}\label{fig:z-frequency}
\end{figure}

\begin{figure}[t!]
\includegraphics[width=.23\textwidth]{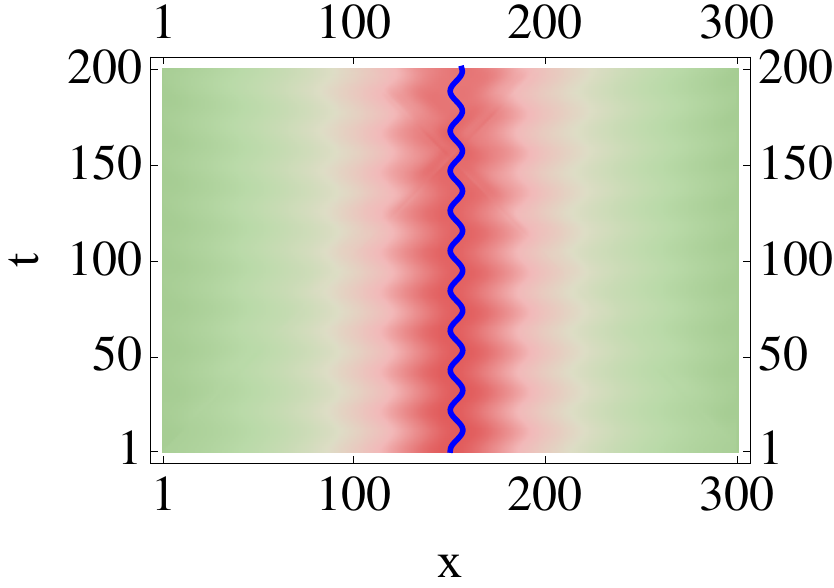}\quad
\includegraphics[width=.23\textwidth]{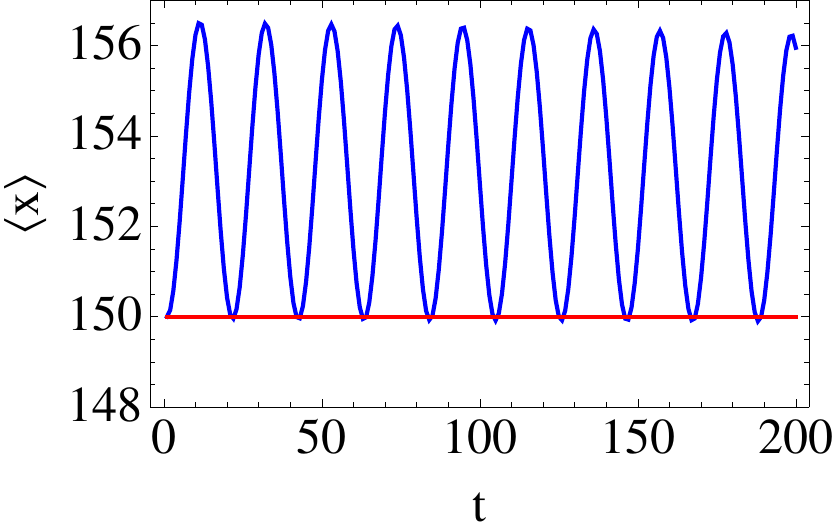}
\includegraphics[width=.23\textwidth]{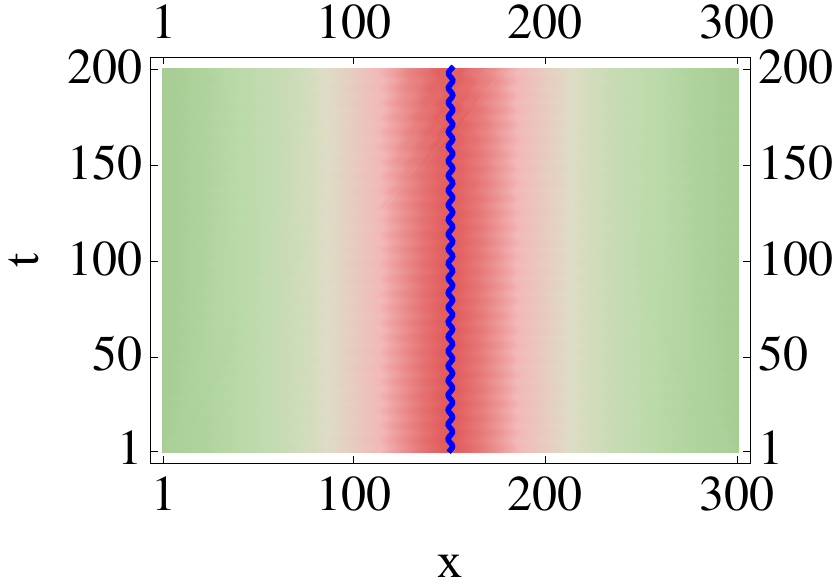}\quad
\includegraphics[width=.23\textwidth]{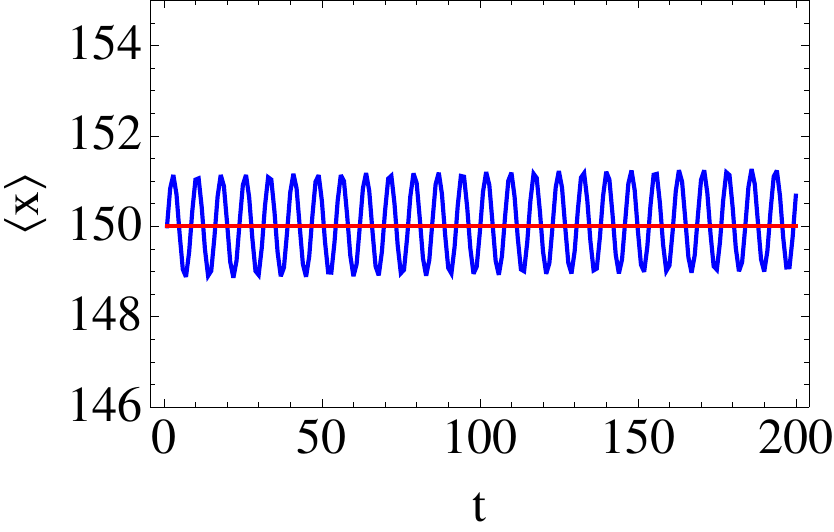}
\includegraphics[width=.23\textwidth]{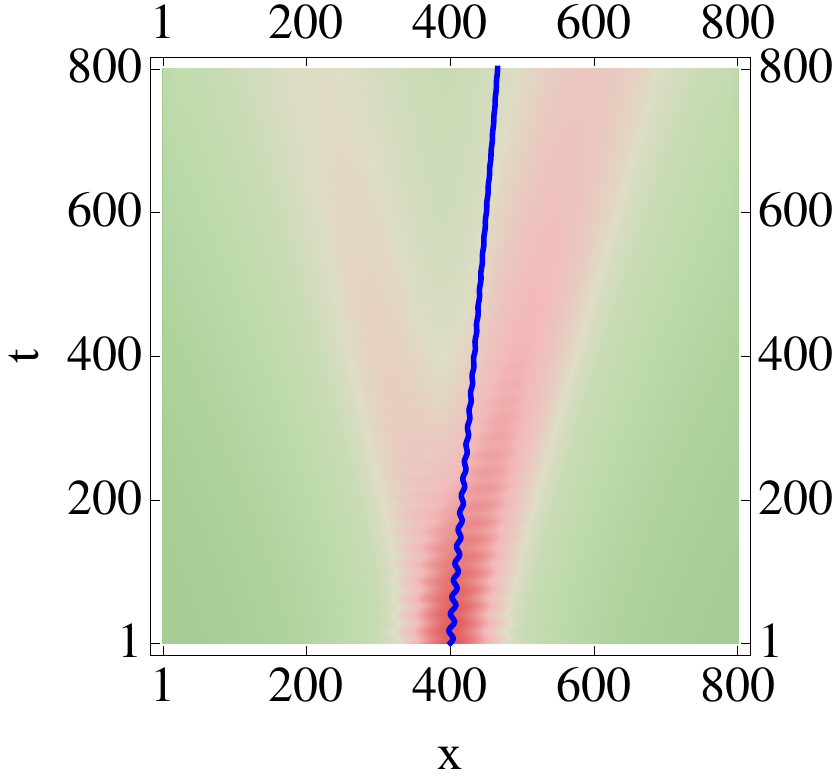}\quad
\includegraphics[width=.23\textwidth]{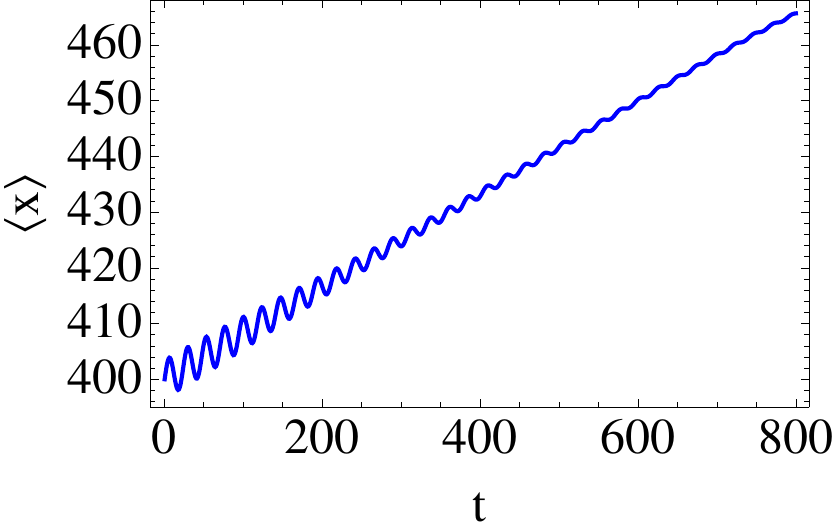}
\caption{ Automaton evolution of a state as in Eq.\eqref{e:states}
  showing the Zitterbewegung of the position expectation value.  {\bf
    Top:} $m=0.15$, $c_+=1/ \sqrt{2}$, $c_-=i/\sqrt{2}$, $k_0=0$, and
  $\sigma=40^{-1}$. The calculated shift and oscillation frequency are
  respectively $\bra{\psi} X(0)+ Z_{X}(0)\ket{\psi}=3.2$ and
  $\omega(0)/\pi=0.05$, accordingly to the simulation.  {\bf Middle:}
  $m=0.15$, $c_+=1/\sqrt{2}$, $c_-=1/\sqrt{2}$, $k_0=0$,
  $\sigma=40^{-1}$.  The calculated shift and oscillation frequency
  are $0$ and $0.13$, respectively.  {\bf Bottom:} $m=0.13$,
  $c_+=\sqrt{2/3}$, $c_-=1/\sqrt{3}$, $k_0=10^{-2}\pi$,
  $\sigma=40^{-1}$.  In this case the particle and antiparticle
  contribution are not balanced and the average position drift
  velocity is thus $\bra{\psi _+} V \ket{\psi_+}+\bra{\psi _-} V
  \ket{\psi_-}=(|c_+|^2-|c_-|^2)v(k_0)=0.08$, corresponding to an
  average position $x_{\psi}^+(800) + x_{\psi}^-(800)=464$ (see
  Eq. \eqref{e:zitterbewegung}). Notice that for $t \to \infty$ the
  term $2 \Re [\bra{\psi_+} {Z}_{{X}}(t) \ket{\psi_-}$, which is
  responsible of the oscillation, goes to
  $0$.}\label{fig:Zitt11-12-21-22}
\end{figure}

In Fig. \ref{fig:Zitt11-12-21-22}  we have considered the evolution of states with particle
and antiparticle components smoothly peaked around some momentum
eigenstate, namely
\begin{align}\label{e:states}
c_+\ket{\psi_+}+c_-\ket{\psi_-},\quad \ket{\psi_\pm}= \int
  \df{k}{\sqrt{2\pi}} g_{k_0}(k) \ket{\pm}_k\ket{k}
\end{align}  
where $c_+^2 \mkern-6mu+\mkern-6mu c_-^2 \mkern-14mu = \mkern-14mu 1$ and $g_{k_0}$  is a Gaussian peaked around the
momentum $k_0$ with width $\sigma$. An easy computation shows that for these states
the shift contribution reduces to   $2\Re[\bra{\psi}X(0)+
Z_{X}(0)\ket{\psi}]=\Im{(c^{*}_+c_-)}/(2\pi)\int_{-\pi}^{\pi}\,\d{k}
|g_{k_0}(k)|^2 z(k)$ with the function 
$z(k)=m\cos\omega(k)/\sin^2\omega(k)$ bounded again by the Compton
wavelength $1/m$ and the oscillation frequency given by $\omega(0)/\pi$
(see also Fig. \ref{fig:z-frequency}).


\section{Evolution with a square potential barrier}\label{s:barrier}

In order to study the scattering with a potential, we modify the automaton adding a position
dependent phase representing a square potential barrier, as in Refs.
\cite{kurzynski2008relativistic,meyer1997quantum}.  We will provide explicitly the transmission $T$ and
reflection $R$ coefficients as functions of the energy and mass of the incident wavepacket and of
the potential barrier's height. We will find a general behavior independently on the regime, namely
on the energy and mass of the incident particle.  Increasing the value of the potential barrier
beyond a certain threshold a transmitted wave reappears and the reflection coefficient starts
decreasing. The width of the $R=1$ region is an increasing function of the mass which is
proportional to the gap between positive and negative frequency eigenvalues of the unitary
evolution.

\begin{figure}[h!]
\includegraphics[width=.30\textwidth]{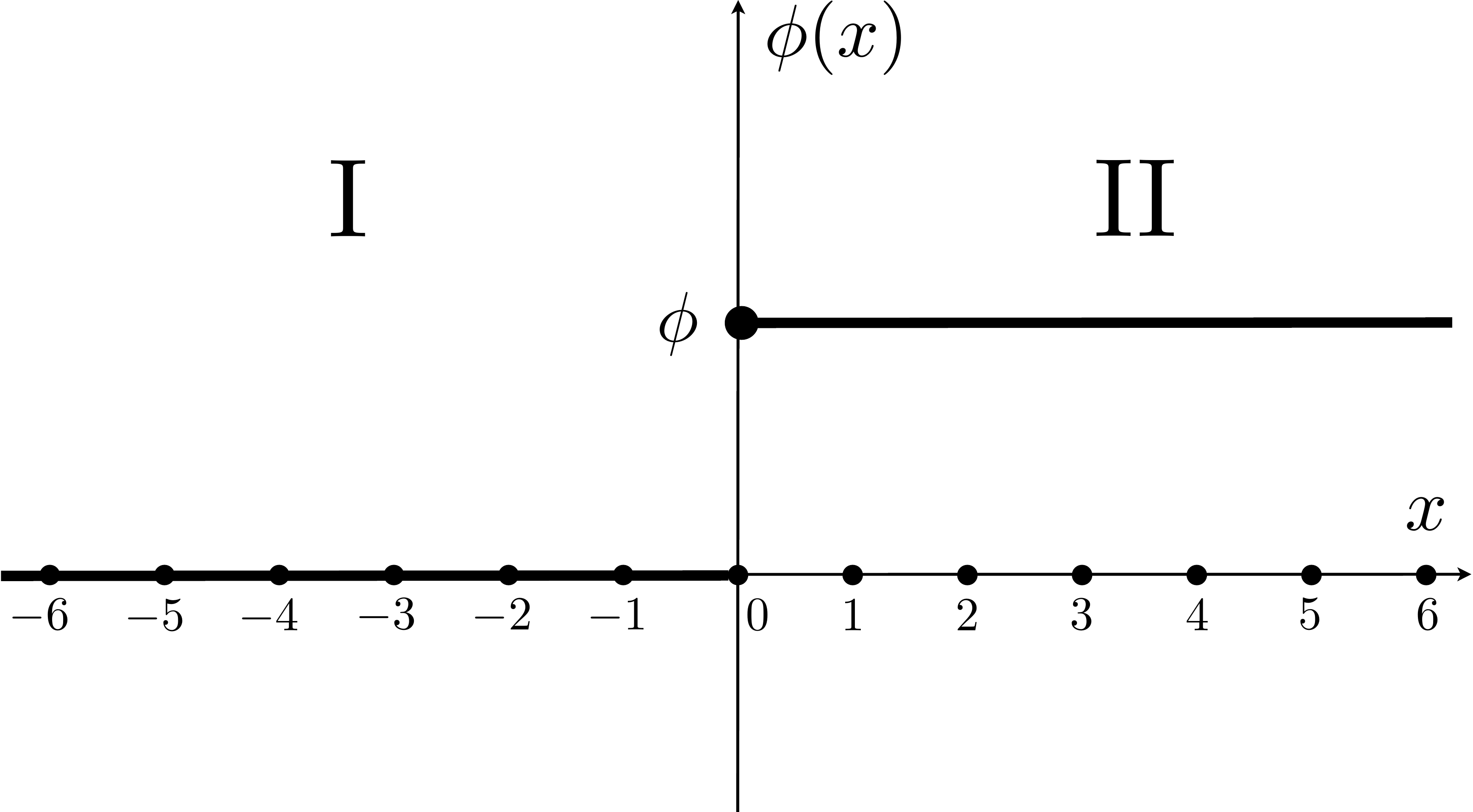}
\caption{Schematic of the potential}\label{fig:potential}
\end{figure}

For a generic potential $\phi(x)$, the unitary evolution becomes
\begin{align*}
  U_\phi := \sum_x e^{-i \phi(x)}
\left(  
\begin{array}{ll}
   n \ketbra{x-1}{x} &   -im \ketbra{x}{x} \\
  -im \ketbra{x}{x} &   n \ketbra{x+1}{x} 
  \end{array}
\right).
\end{align*}
We will analyze the simple case $\phi(x) := \phi \, \theta(x)$
($\theta(x)$ is the Heaviside step function) that is a potential step  which is $0$
for $x <0 $ (region $\mathrm{I}$) and has a constant value $\phi \in
[0, 2\pi]$ for
$x \geq 0$
(region $\mathrm{II}$)
as illustrated in Fig. \ref{fig:potential}.

\begin{figure}[b!]
\includegraphics[width=.23\textwidth]{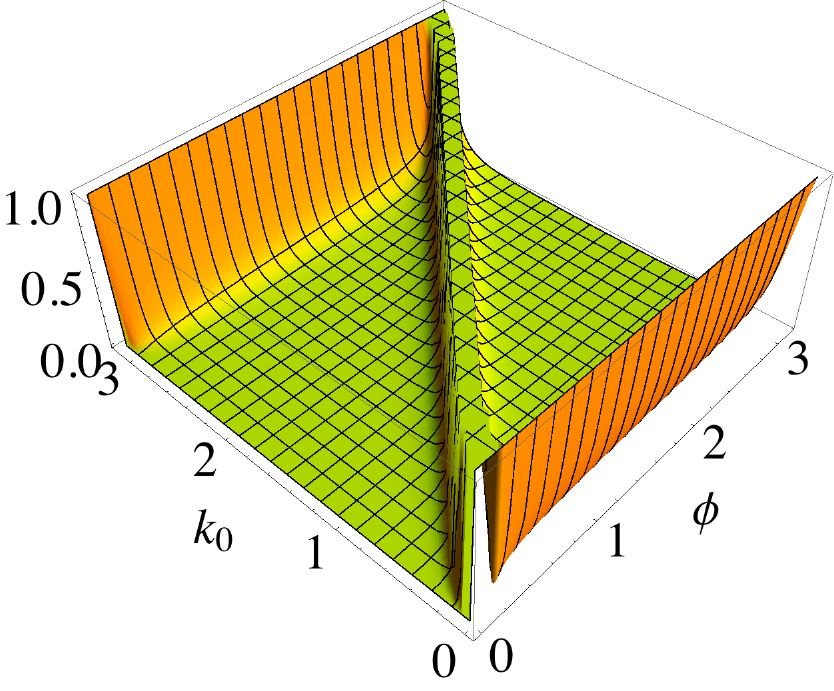}\quad
\includegraphics[width=.23\textwidth]{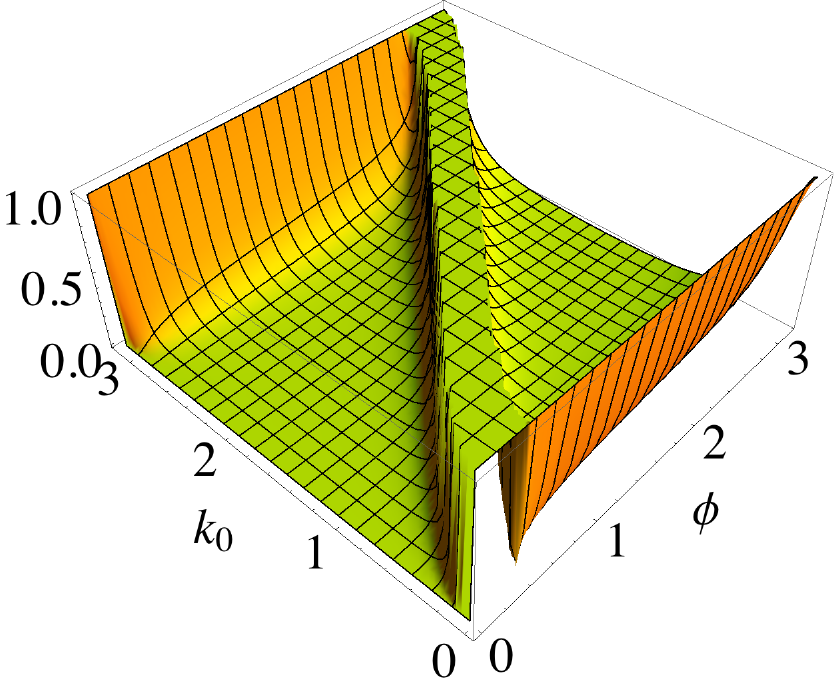}
\includegraphics[width=.23\textwidth]{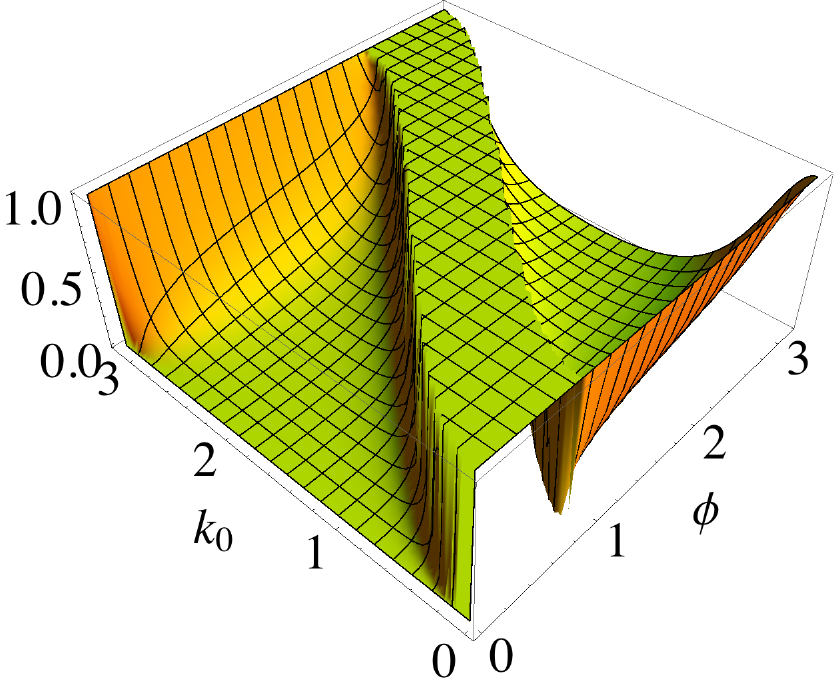}\quad
\includegraphics[width=.23\textwidth]{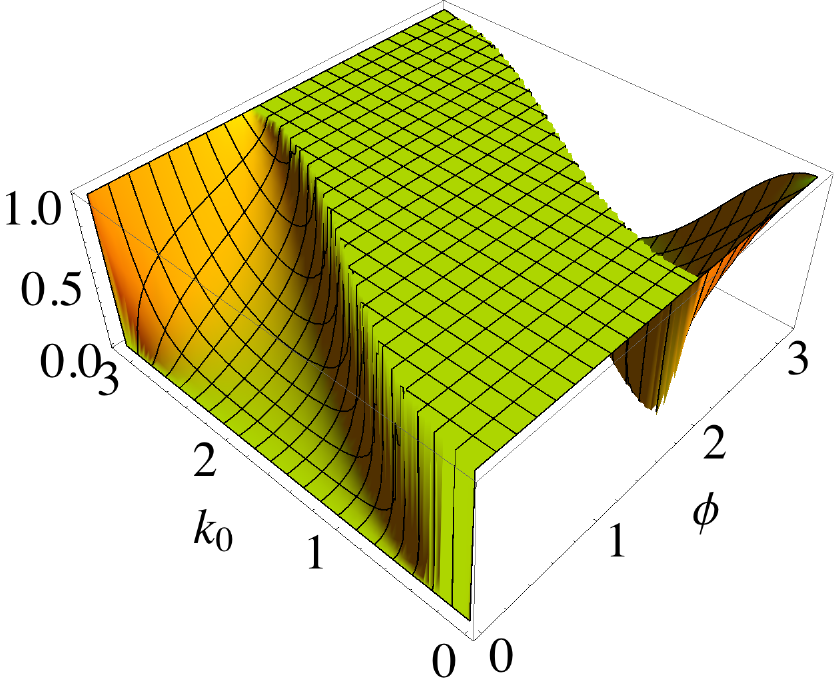}
\caption{Reflection coefficient as a function of the potential barrier
  height $\phi$ and of the momentum $k$ of the incident particle
  state. From the top-left to the bottom-right the reflection
  coefficient is depicted for different values of the mass: $m=0.1,\,0.2,\,0.4,\,0.8$.}\label{fig:Reflection}
\end{figure}
\begin{figure}[b!]
\includegraphics[width=.23\textwidth]{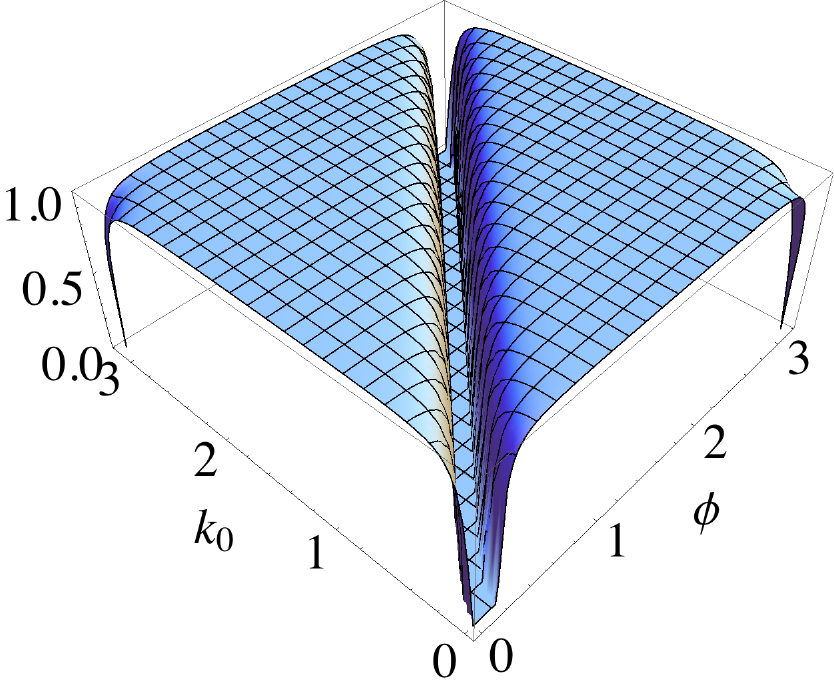}\quad
\includegraphics[width=.23\textwidth]{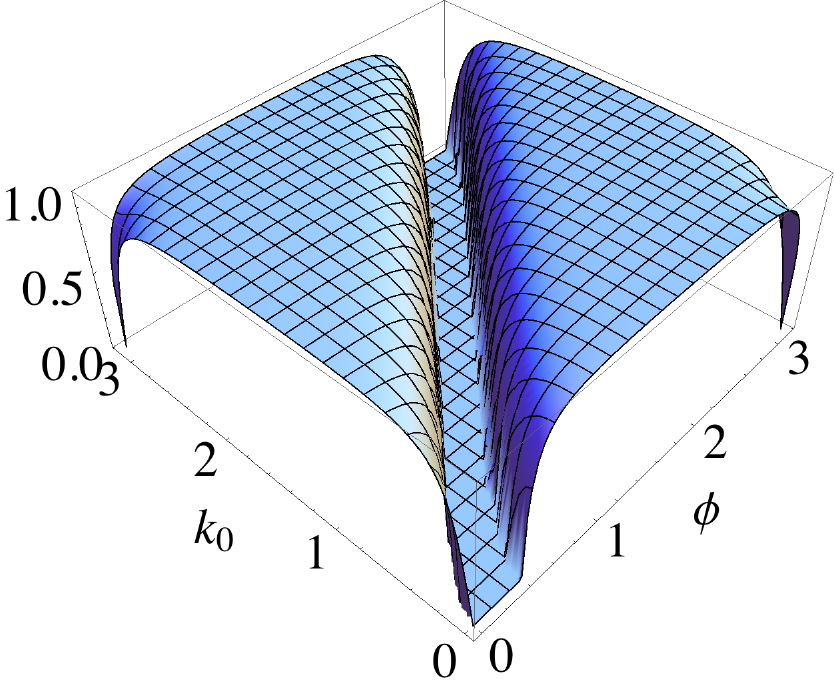}
\includegraphics[width=.23\textwidth]{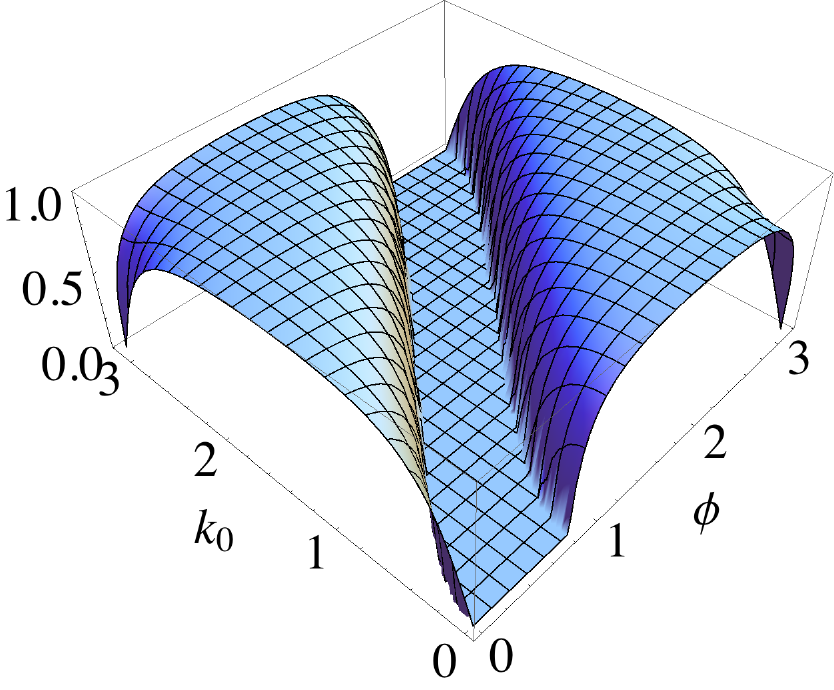}\quad
\includegraphics[width=.23\textwidth]{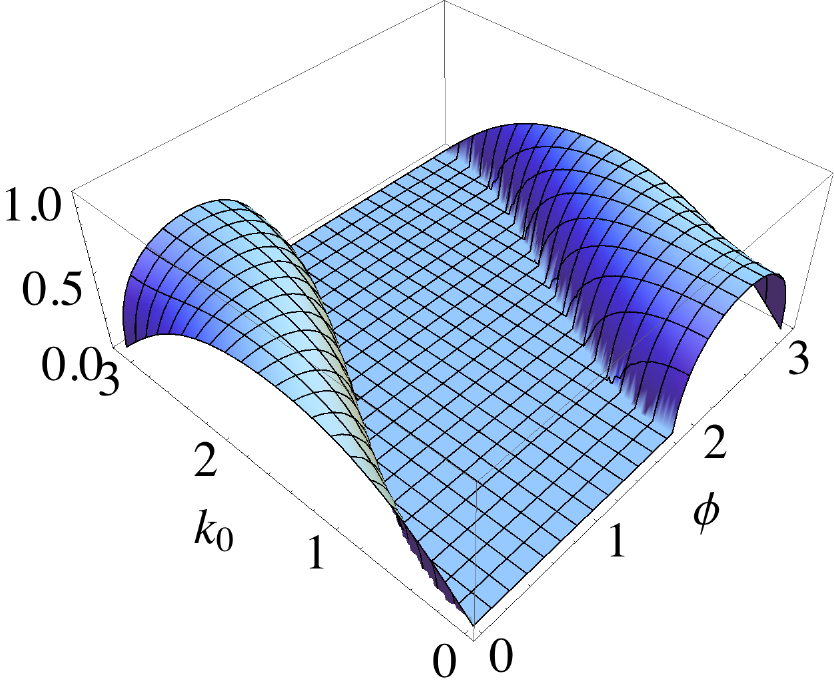}
\caption{Group velocity of the transmitted wave packet as a function
  of the potential barrier height $\phi$ and of the momentum $k$ of the
  incident particle state. From the top-left to the bottom-right the
  transmitted group velocity for different values of the mass:
  $m=0.1,\,0.2,\,0.4,\,0.8$.}\label{fig:Momentum}
\end{figure}

Let us now study the eigenvector of $U_\phi$ 
of the form
\begin{align*}
    &\ket{\Phi_k} =     \Pi_\mathrm{I} \ket{+}_k\ket{k}  +\Pi_\mathrm{I} \beta_k
    \ket{+}_{-k}\ket{k}  +  \gamma_k \Pi_\mathrm{II}\ket{+}_{k'}\ket{k'}    \\
    &\Pi_\mathrm{I} := \sum _{x<0} I \otimes \ketbra{x}{x} \qquad
\Pi_\mathrm{II} := \sum _{x \geq 0} I \otimes \ketbra{x}{x} 
 \end{align*}
where 
$\beta_k$, $\gamma_k$ and $k'$ are  functions of $k$.
The condition that $\ket{\Phi_k}$ is genuinely an eigenstate
of $U_\phi$, i.e. $  U_\phi \ket{\Phi_k} = e^{-i \omega(k)}
\ket{\Phi_k}$,  implies that 
\begin{align}
&\omega(k') = \omega(k) -\phi  \label{eq:kprime}\\
&\beta_k=\frac{e^{-ik}\sqrt{(1+ v)(1-v^\prime)}-e^{-ik^\prime}\sqrt{(1-
    v)(1+v^\prime)}}{-e^{ik}\sqrt{(1- v)(1-
    v^\prime)}+e^{-ik^\prime}\sqrt{(1+v)(1+v^\prime)}} \nonumber \\
&\nonumber
\gamma_k =\frac{2e^{i\xi} (v\cos{k}-i\sin{k})}{-e^{ik}\sqrt{(1- v)(1- v^\prime))}+e^{-ik^\prime}\sqrt{(1+v)(1+v^\prime)}}
\end{align}
with $v:=v(k)$ and $v':=v(k')$ the group velocities of the incident and transmitted wave.
Let us now consider the superposition
\begin{align*}
\ket{\Psi(0)}:=  \int \df{k}{\sqrt{2\pi}} g_{k_0}(k) \ket{\Phi_k}
\end{align*}
where $g_{k_0}(k)$ is a function in $C_0^{\infty}[-\pi, \pi]$  
which we assumed to be smoothly peaked around $k_0$. 
The state at time $t$ is then
\begin{align*}
\ket{\Psi(t)}:=  \int \df{k}{\sqrt{2\pi}} g_{k_0}(k) e^{-i \omega(k)t} \ket{\Phi_k}
\end{align*}
and one can verify that for $t \ll 0$ the state is negligible in region $\mathrm{II}$ while the
only appreciable contribution in region $\mathrm{II}$ comes from the term
$e^{i k_0 x}$ which describes a wavepacket  that moves at group velocity
$v(k_0)$ and hits the barrier form the left.
When $t \gg 0$ the state can be approximated by a  superposition of a reflected
and a transmitted wavepacket as follows
\begin{align*}
  \begin{split}
    \ket{\Psi(t)} \xrightarrow{ t \gg 0} \beta(k_0) \int \df{k}{\sqrt{2\pi}}  g_{k_0}(k) e^{-i \omega(k)t}  \ket{+}_{-k}\ket{k} +\\
 {}+ \tilde{\gamma}(k_0) e^{-i \phi t} \int \df{k}{\sqrt{2\pi}}  \tilde{g}_{k'_0}(k') e^{-i
    \omega(k')t} \ket{+}_{k'}\ket{k'} 
  \end{split}
\end{align*}
where we defined
\begin{align*}
&k'_0 \mbox{ s.t. }    \omega(k'_0) = \omega(k_0) -\phi, \\
&\tilde{\gamma}(k_0) := {\gamma}(k_0)
\sqrt{\frac{v(k'_0)}{v(k_0)}}, \qquad
\tilde{g}_{k'_0}(k') =
\sqrt{\frac{v(k'_0)}{v(k_0)}}{g}_{k'_0}(k') 
\end{align*}
(one can check $ \int \df{k}{\sqrt{2\pi}} 
|\tilde{g}_{k'_0}(k')|^2 = 1$), whose group velocities are
$-v(k_0)$ for the reflected wave packet and $v(k^\prime_0)$ for the
transmitted wave packet (see Fig. \ref{fig:Momentum}).

The probability of finding the particle in the reflected wavepacket is $R = |\beta(k_0)|^2$
(reflection coefficient) while the probability of finding the particle in the transmitted
wavepacket is $ T= |\tilde{\gamma}(k_0)|^2$ (trasmission coefficient). The consistency of the result
can be verified by checking that $R+T = 1$. For $k \ll m \ll 1 $ (Schr\"oedinger regime) we recover
the usual reflection and transmission coefficient for the Schr\"oedinger equation with a potential
step.  In Fig. \ref{fig:Reflection} we plot the reflection coefficient $R$ as a function of $\phi$
and $k$ for different values of the mass $m$. Clearly when $\phi=0$ we
have $R = 0$ and increasing $\phi$ while fixing $k$ the value increases up to $R=1$.  One notice that when $
\omega(k) - \arccos(n) < \phi < \omega(k) + \arccos(n)$ Eq.~\eqref{eq:kprime} has solution
for imaginary $k'$ which implies an exponential damping of the transmitted wave and pure reflection.
By further increasing the value of $\phi$ beyond the threshold $\omega(k) + \arccos(n)$,
Eq.~\eqref{eq:kprime} have solution for real $k'$ and negative $\omega(k')$, and then a
transmitted wave reappears and the reflection coefficient decreases. This is the so called ``Klein
paradox'' which is originated by the presence of positive and negative frequency eigenvalues of the
unitary evolution.  The width of the $R=1$ region is an increasing function of the mass equal to
$2\arccos(n)$ which is the gap between positive and negative frequency solution solutions (see
Fig. \ref{fig:disp} ).
\begin{figure}[t!]
\includegraphics[width=.23\textwidth]{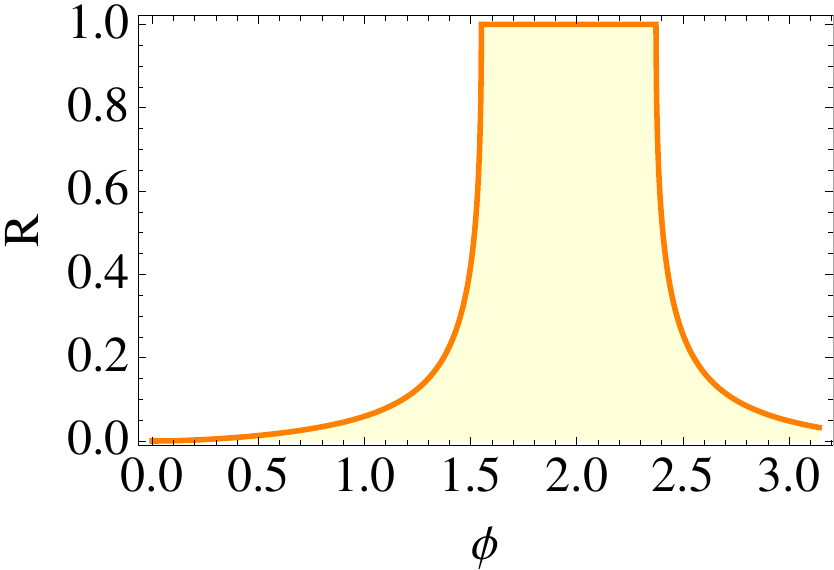}\quad
\includegraphics[width=.23\textwidth]{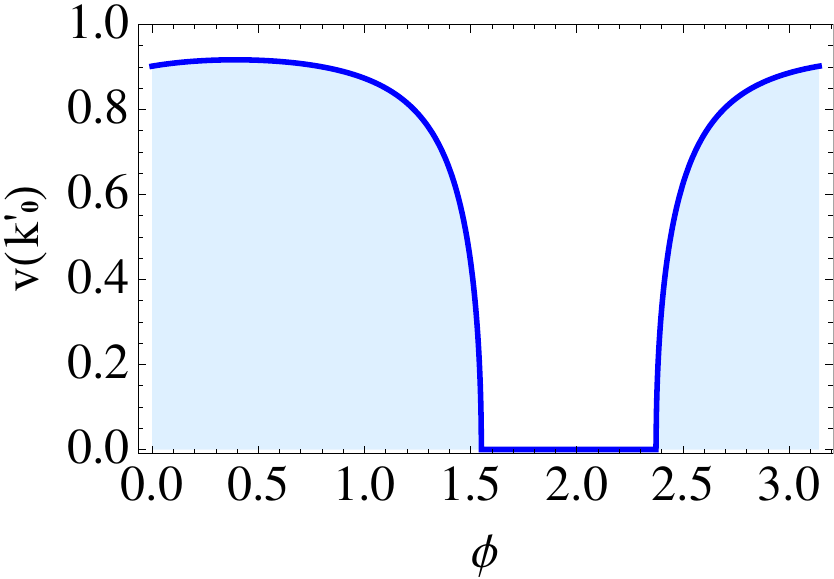}
\caption{Reflection coefficient for $m=0.4$ and momentum of the
  incident particle $k_0=2$ as a function of the potential barrier
  height $\phi$ (section of plots in
  Figs. \ref{fig:Reflection}-\ref{fig:Momentum} for
  $m=0.4$, $k_0=2$).}\label{fig:Section}
\end{figure}
\begin{figure}[t!]
\includegraphics[width=.23\textwidth]{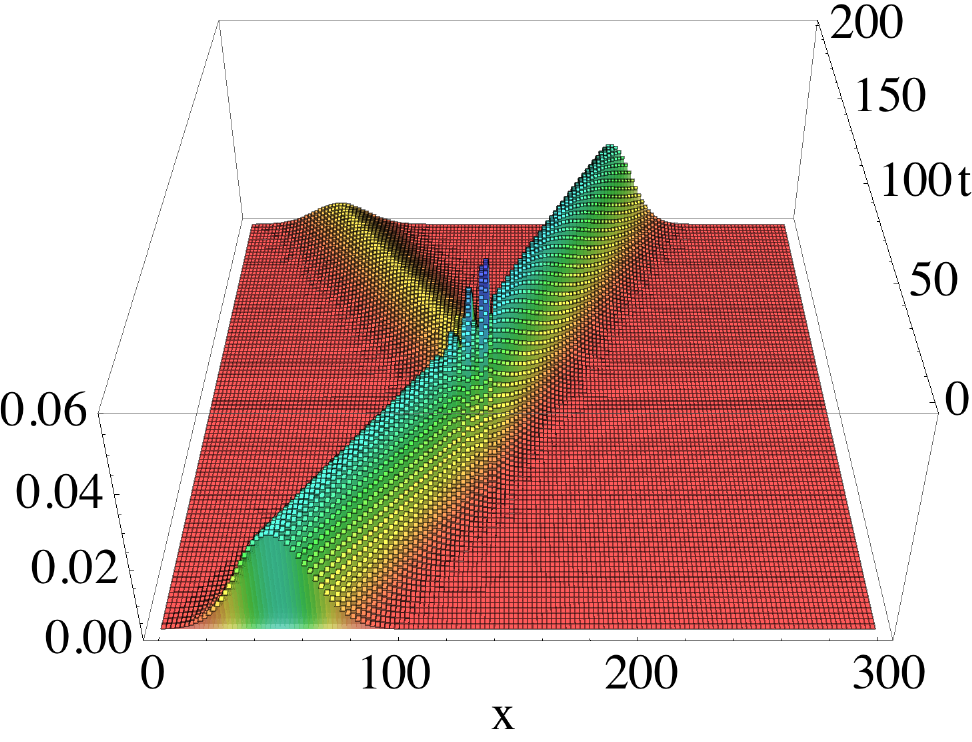}\quad
\includegraphics[width=.23\textwidth]{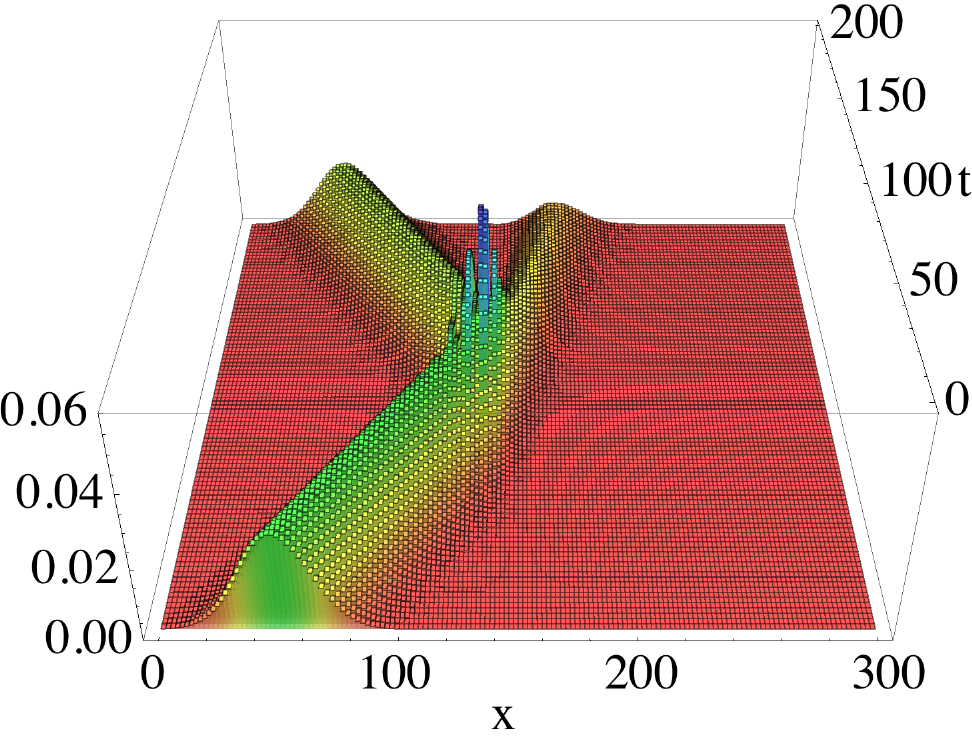}
\includegraphics[width=.23\textwidth]{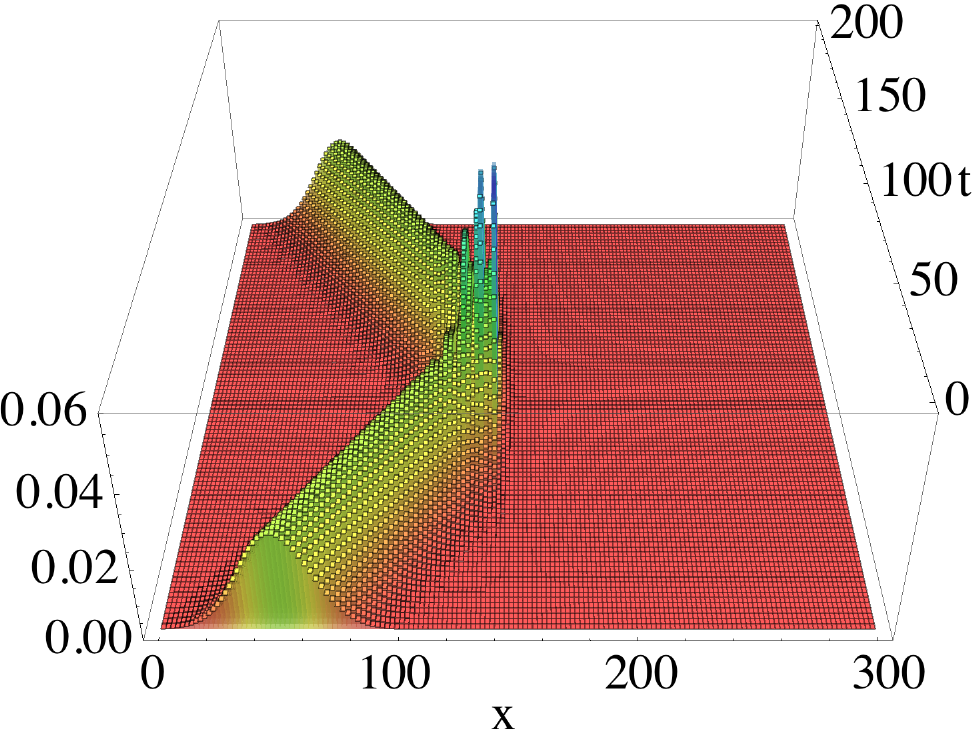}\quad
\includegraphics[width=.23\textwidth]{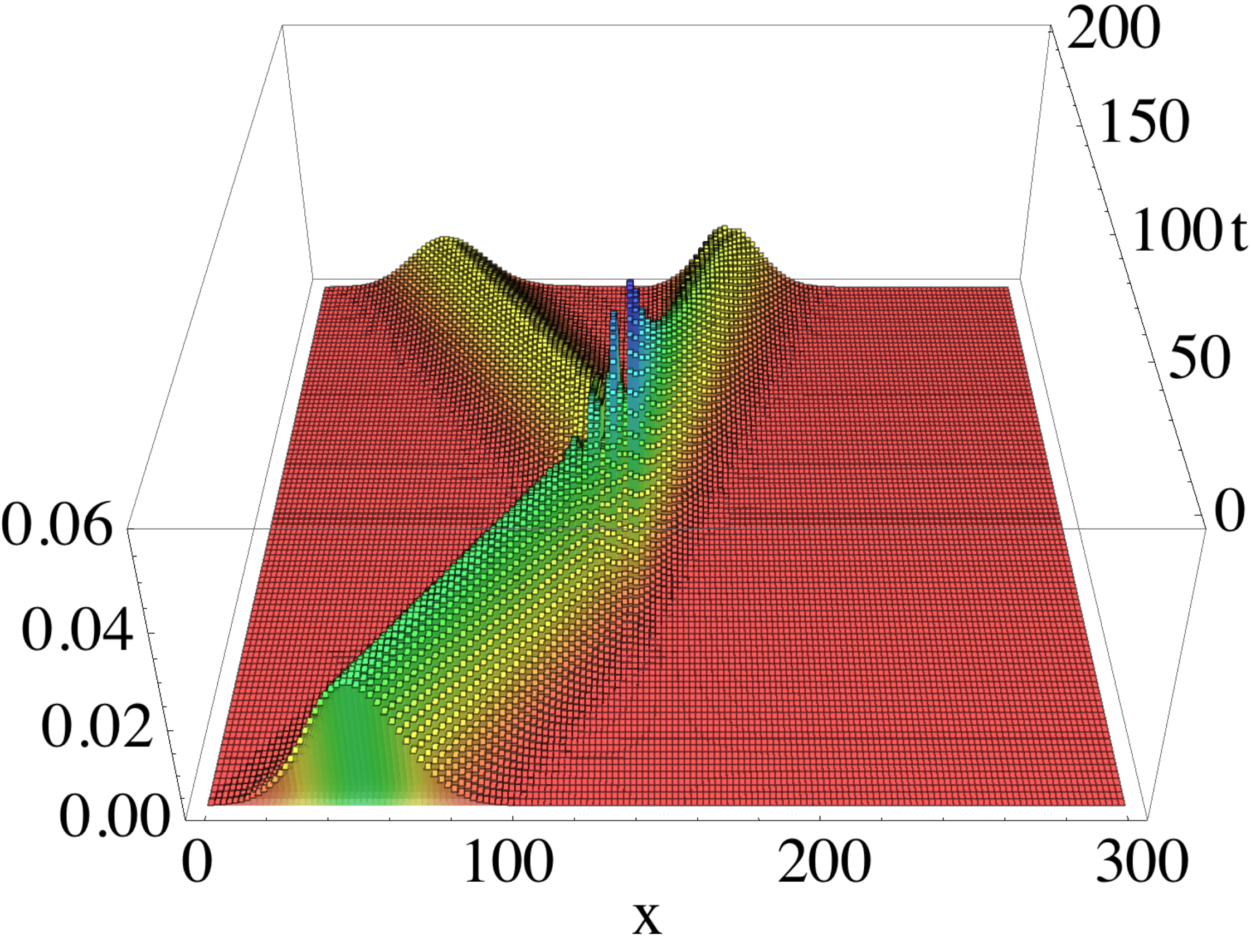}
\caption{Simulations of the Dirac automaton evolution with a square
  potential barrier. Here the automaton mass is $m=0.2$ while the
  barrier turns on at $x=140$. In the simulation the incident state
  is a smooth state of the form $\ket{\psi(0)}=  \int
  \df{k}{\sqrt{2\pi}} g_{k_0}(k) \ket{+}_k$ peaked around the positive energy eigenstate
  $\ket{+}_{k_0}$ with $k_0=2$ and
  with $g_{k_0}$ a Gaussian having width
  $\sigma=15^{-1}$. The incident group velocity is $v(k_0)=0.90$. The
  simulation is run for four increasing values of the potential $\phi$. 
  {\bf  Top-Left:} 
 Potential barrier height $\phi=1.42$, reflection coefficient
  $R=0.25$, velocity of the transmitted particle $v(k^\prime_0)=0.63$.  
  {\bf Top-Right:} $\phi=1.55$, $R=0.75$, $v(k^\prime_0)=0.1$.
  {\bf Bottom-Left:} $\phi=2$, $R=0.1$, $v(k^\prime_0)=0$.  
  {\bf Bottom-Right:} $\phi=2.4$, $R=0.50$, $v(k^\prime_0)=0.33$.} \label{fig:Simulation}
\end{figure}

In Fig. \ref{fig:Section} we plot the reflection $R$ coefficient and the transmitted wave velocity
group $v(k_0')$ as a function of the potential barrier height $\phi$ with the incident wave packet
having $k_0=2$ and $m=0.4$. From the figure it is clear that after a plateau with $R=1$ the
reflection coefficient starts decreasing for higher potentials. In Fig. \ref{fig:Simulation} we show
the scattering simulation for four increasing values of the potential, say
$\phi=1.42,\,1.55,\,2,\,2.4$ (see the caption to figure for the details).

\section{Conclusions}\label{s:concl}
In this paper we studied the dynamics of the quantum cellular automaton of Refs.
\cite{darianopla,BDTqcaI}, which gives the Dirac dynamics as emergent in the limit of small
wavevectors. We presented computer simulations and analytical evaluations, focusing on typical
features of the Dirac dynamics, in particular the Zitterbewegung and the scattering from potential.
Our automaton covers all regimes of masses and energy-momenta, beyond the same validity range of the
Dirac equation, with the possibility of considering arbitrary input states, enabling to investigate
and visualize a wide range of fundamental processes. This facts, in addition to the discreteness of
the automaton, makes of it the ideal theoretical counterpart for the experimental simulators in the
literature. A similar quantum cellular automaton can be also developed in two dimensions \cite{DP},
corresponding to the graphene as quantum simulator.

\appendix

\section{Bound of the oscillating term and its asymptotic behavior}\label{a:zitterbewegung}

Here we provide an upper bound for the oscillating term $x_{\psi}^{\rm{int}}(t)$ in Eq.
(\ref{e:zitterbewegung}) in the position operator evolution derived in Section
\ref{sec:zitterbewegung} and we derive its behaviour for very long
time steps. The jittering of the position expectation
 value is caused by the operator ${Z}_{{X}}(t)$ which
in the base diagonalizing the automaton
 Hamiltonian $H$ \eqref{eq:hamiltonian} can be written as
\begin{align}\nonumber
Z_X(t)=\int_{-\pi}^{\pi}\d k e^{2i\omega(k) \sigma_z
  t}Z_X(k)\otimes\ketbra{k}{k},\\\nonumber
Z_X(k)=z(k)\sigma_2,\qquad 
z(k)=\frac{m\cos{\omega(k)}}{2\sin^2{\omega}(k)}
\end{align}
with $z(k)\in L^1(-\pi,\pi)$ for any $m\neq 0$.
By defining 
\begin{align}\nonumber
 \ket{\psi_\pm}= \int_{-\pi}^{\pi}\;
  \df{k}{\sqrt{2\pi}} g_{\pm}(k) \ket{\pm}_k\ket{k},\qquad g_{\pm}(k)\in C_0^{\infty}[-\pi, \pi]
\end{align} 
we have 
\begin{align}\nonumber
2\Re[\bra{\psi_+}Z_{X}{(t)}\ket{\psi_-}]=\int_{-\pi}^{\pi} \df{k}{\pi} z(k)
\Re{\left[i g^*_+(k) g_-(k) e^{2i\omega(k) t}\right]}
\end{align}
Since, for any $m\neq 0$,  $\omega(k)$ has three stationary
points in $k=0,\,\pm\pi$ ($\omega^{(1)}(0)=\omega^{(1)}(\pm\pi)=0$ and
$\omega^{(1)}(k)\neq 0$ elsewhere in the closed interval
$k\in[-\pi,\pi]$, with $\omega^{(2)}(0),\omega^{(2)}(\pm\pi)\neq
0$), the stationary phase approximation gives 
\begin{multline}\label{e:asymptotic}\nonumber
2\Re[\bra{\psi_+}Z_{X}{(t)}\ket{\psi_-}]\xrightarrow{ t \gg 0}\\\nonumber
\qquad\qquad\sum_{k=0,\pm\pi} z(k)\Re{\left[ig^*_+(k) g_-(k)
e^{2i\omega(k) t}\sqrt{\frac{i}{\pi\omega^{(2)}(k)t}}\;\right]}
\end{multline}
showing that the term $2 \Re [\bra{\psi_+} {Z}_{{X}}(t)
\ket{\psi_-}]$, goes to $0$
as $1/\sqrt{t}$. 

In order to find an upper bound for $x_{\psi}^{\rm{int}}(t)$ notice
that 
\begin{align}\nonumber
  | x_{\psi}^{\rm{int}}(t)|&\leq 2 | \bra{\psi_+} X(0) - {Z}_{{X}}(0)
  + {Z}_{{X}}(t) \ket{\psi_-}|\\\nonumber 
&\leq 2(|\bra{\psi_+} X(0)
  \ket{\psi_-}| + | {Z}_{{X}}(0)| + |{Z}_{{X}}(t)|)
 \end{align}
and, according to the expression of $Z_X(k)$, we get
\begin{align}\nonumber
   | {Z}_{{X}}(0)| + |{Z}_{{X}}(t)| \leq 2| {Z}_{{X}}(0)|\\\nonumber
   | {Z}_{{X}}(0)|\leq \max_{k\in[-\pi,\pi]} |z(k)|=z(0)=\frac{\sqrt{1-m^2}}{2m}.
\end{align}
Now defining the  $C_0^{\infty}[-\pi, \pi]$ test function $\varphi(k,k')=g^*_+(k) g_-(k')
\bra{+}{}_k \ket{-}_{k'}$, we have
\begin{align}\nonumber
&|\bra{\psi_+} X(0) \ket{\psi_-}| =\\\nonumber 
&\left|\BraKet{
\frac
{\;\dd\delta(k-k')}{\;\dd (k-k')}
}{\varphi(k,k')}\right|=
\left|\BraKet{\delta(k-k')}
{
\frac
{\;\dd\varphi(k,k')}{\;\dd (k-k')}
}\right|=\\\nonumber
&\left|\int_{-\pi}^{\pi} \;\df{k}{2\pi}\, \d k' \delta(k-k') g^*_+(k) g_-(k')
\frac{\;\dd }{\;\dd (k-k')}\bra{+}{}_k \ket{-}_{k'}\right|=\\\nonumber
&\left|\int_{-\pi}^{\pi}\; \df{k}{2\pi}  g^*_+(k) g_-(k) f(k) \right|
\leq \max_{k\in[-\pi,\pi]} |f(k)|=f(0)\\\nonumber
&f(k):=\frac{n}{\sin^2{\omega}},\qquad f(0)= \frac{\sqrt{1-m^2}}{m^2}.
\end{align}
which finally gives
\begin{align}
 | x_{\psi}^{\rm{int}}(t)| \leq \frac{2}{m}+\frac{2}{m^2}.
\end{align}

\bibliography{bibliography}

\end{document}